\documentstyle[12pt,psfig]{article}
  
\begin{document}
\baselineskip=0.8cm
\centerline{\bf Vigorous star formation hidden by dust in a galaxy at $z=1.4$} 

\bigskip
\bigskip

\centerline{\bf Andrea Cimatti$^{1}$, Paola Andreani$^{2}$,} 

\centerline{\bf Huub R\"ottgering$^{3}$, Remo Tilanus$^{4,5}$}

\bigskip

\centerline{$^{1}$ {\it Osservatorio Astrofisico di Arcetri,
Largo E. Fermi 5, I-50125, Firenze, Italy}}

\centerline{$^{2}$ {\it Dipartimento di Astronomia, vicolo dell'Osservatorio 5,
I-35122 Padova, Italy}}

\centerline{$^{3}$ {\it Sterrewacht, Postbus 9513, Leiden, 2300 RA, 
Netherlands}} 

\centerline{$^{4}$ {\it Joint Astronomy Centre, 660 N. A'ohoku Place, Hilo, 
Hawaii 96720, USA}}

\centerline{$^{5}$ {\it Netherlands Foundation for Research in Astronomy,
Dwingeloo, The Netherlands}}

\bigskip
\bigskip
\centerline{\sl To be published in Nature, 30 April 1998 issue}
\bigskip
\bigskip

{\bf Near-infrared surveys have revealed a substantial population 
of enigmatic faint galaxies with extremely red optical-to-near-infrared 
colours and with a sky surface density comparable to that of faint 
quasars \cite{hu}. There are two scenarios for these extreme 
colours: (i) these distant galaxies have formed virtually all their 
stars at very high redshifts and, due to the absence of recently formed 
stars, the colours are extremely red \cite{dunlop} and (ii) these 
distant galaxies contain large amounts of dust, severely reddening the rest-frame 
UV--optical spectrum. HR10 ($z = 1.44$, \cite{hu,graham}) is considered the
archetype of the extremely red galaxies.  Here we report the detection
of the continuum emission from HR10 at 850$\mu$m and at 1250$\mu$m,
demonstrating that HR10 is a very dusty galaxy undergoing a major episode 
of star formation. Our result provides a clear example 
of a high-redshift galaxy where the star formation rate inferred from the 
ultraviolet luminosity would be underestimated by a factor up to 1000, 
and shows that great caution should be used to infer the global 
star formation history of the Universe from optical observations only.
}

The current knowledge of the global star formation activity in the 
Universe is mostly based on the UV continuum luminosity of high-$z$ 
galaxies \cite{madau}. However, a crucial phase in the evolution of 
a galaxy might be a vigorous starburst producing copious amounts of
luminous stars. In this stage, most of the UV stellar
radiation would be absorbed by the dust grains and re-emitted in 
the far-infrared ($FIR$), making the galaxy dark at UV and optical
wavelengths. A cosmic background has been first detected at 
$\lambda_{obs}\sim$200$\mu$m--2{\it mm} \cite{puget} and then 
interpreted by \cite{guid}. In addition, a cosmic background
at 140$\mu$m and 240$\mu$m has been recently found by \cite{schlegel}.
The presence of such a background suggests the existence of a population 
of distant dusty galaxies. 
Several models predict the number of dusty galaxies at 
high-$z$ \cite{blain,fra,zepf,guid}. However, the presently known dusty 
galaxies at $z>1$ are limited to a handful of extremely rare objects such 
as radio galaxies, quasars and $IRAS$-selected luminous galaxies
(\cite{hughes,sand} and references therein), and the existence of a 
general field population of dusty galaxies has not been established yet.
Recent surveys for field galaxies found faint ($R>$24) and very red 
objects with colours typically in the range of 7$<R-K<$8 
\cite{hu, thompson}. Their sky surface density, $\approx$0.01-0.02 
arcmin$^{-2}$ at $K \leq 20$ \cite{hu, thompson}, is comparable to
that of faint quasars \cite{hu}, but about one order of magnitude 
less than that of field galaxies with $z>3$ \cite{steid}.

In order to test if a fraction of these galaxies have these extremely 
red colours because of strong dust reddening, we started a programme 
aimed at detecting their (sub)-mm dust thermal emission. Our first 
target was HR10, one of the reddest
galaxies known to date, with $I=24.9$, $R-K\sim8$, $I-K=6.5$ \cite{hu} 
and the only with a measured redshift ($z=1.44$ \cite{graham}). Its UV-optical
spectral energy distribution (SED) suggested the presence of dust
reddening \cite{graham,cima}. HR10 was 
observed and detected with the Institut de Radioastronomie Millimetrique 
($IRAM$) 30m antenna at 1250$\mu$m as well as with the James Clerk Maxwell 
Telescope ($JCMT$)\footnote{The JCMT is operated by the Joint Astronomy 
Centre on behalf of the Particle Physics and Astronomy Research Council 
of the United Kingdom, the Netherlands Organisation for Scientific 
Research, and the National Research Council of Canada} 15m telescope at 
850$\mu$m in March and September 1997 respectively. The details of the 
observations and data reduction can be found in the legend of Figure 1. 
Throughout the paper we assume $H_0=50$ kms$^{-1}$ Mpc$^{-1}$, $q_0=0.5$, 
and we define $h_{50}=H_0/50$.

Figure 1 shows the SED of HR10. Synchrotron radiation emission as the 
source of the sub-mm fluxes can be ruled out because the radio fluxes 
\cite{graham,jones} constrain the extrapolated synchrotron flux at 
1250$\mu$m to be $<$0.34 mJy, more than an order of magnitude less 
than the observed one. We therefore interpret the continuum emission 
observed at 850$\mu$m and 1250$\mu$m as due to thermal dust emission. 
Using the $ISO$ upper limit at 175$\mu$m \cite{ivi}, and for dust 
emissivity indices $\beta=1-2$ \cite{whittet}, the temperature is 
found to be in a range 18$<T_{d}<$45 K. These temperatures fall into 
the range of those of active and star-forming galaxies (\cite{andre,
sand,hughes} and references therein). The total dust mass can be 
estimated as \cite{hildebrand}: $M_{d} = S_{\nu_{obs}}D^{2}_{L}/[(1+z)\kappa_{d}
(\nu_{rest})B(\nu_{rest},T_{d})]$, where $\nu_{obs}$ and $\nu_{rest}$
are, respectively, the observed and rest-frame frequencies, $S_{\nu_{obs}}$
is the observed flux density at $\nu_{obs}$, $D_{L}$ is the luminosity
distance, $B$ is the black-body Planck function, $T_{d}$ is the dust 
temperature and $\kappa_{d}= 0.67(\nu_{rest}/250\,{\rm GHz})^{\beta}$\,
cm$^{2}$\, g$^{-1}$ is the adopted mass absorption coefficient 
\cite{hughes}. In the two extreme cases considered in Figure 1, the 
total dust mass is $M_{d} = 7.3 \times 10^{8}$ $h_{50}^{-2}$ M$_{\odot}$
and $M_{d} = 3.3 \times 10^{9}$ $h_{50}^{-2}$ M$_{\odot}$ for 
$T_{d}$=45 K and $T_{d}$=18 K respectively. The dust mass
increases or decreases of a factor of about 2 if $\beta$=1
or $\beta$=2 respectively. If we assume $M_{H_2}/M_{d}$=100
\cite{solomon}, the mass of molecular hydrogen is of the order
of $M_{H_2} \approx 10^{11}$ $h_{50}^{-2}$ M$_{\odot}$.
The total $FIR$ luminosity is estimated by integrating the grey-body
curves in the range $\lambda_{rest} =$10-2000$\mu$m, $L_{FIR,rest}=4 
\pi D^{2}_{L} (1+z)^{2} \int S_{\nu,obs} d\nu$, where $D_{\rm L}$ is 
the luminosity distance and $S_{\nu,obs}$ is the flux density.
We find that $L_{FIR,rest} = 3.8 \times 10^{12}$ $h_{50}^{-2}$ 
L$_{\odot}$ and $L_{FIR,rest} = 1.5 \times 10^{13}$ $h_{50}^{-2}$ 
L$_{\odot}$ for $T_{d}$=18 K and $T_{d}$=45 K respectively. 

At low redshifts, HR10 is comparable to the ultra-luminous infrared
galaxies (ULIGs) which are dusty systems with $L_{FIR}>10^{12}$
L$_{\odot}$ and typically $T_{d}\sim$ 30-60 K \cite{sand}, but with dust 
masses $M_{d} \sim 10^{7-8}$ M$_{\odot}$ lower than those inferred for HR10
(\cite{sand} and references therein). At high redshifts, HR10 can be 
compared to $IRAS$10214+4724, an ULIG at $z=2.286$ \cite{rr}
whose properties are severely affected by gravitational lensing.
If the correction for lensing magnification is taken into account
($\approx 30 \times$ in the infrared \cite{eisen}), $IRAS$10214+4724 has 
$L_{FIR}= 4.4 \times 10^{12}$ $h_{50}^{-2}$ L$_{\odot}$ and $M_{d} = 1.1 \times 
10^{8}$ $h_{50}^{-2}$ M$_{\odot}$ comparable to those of HR10, but it has
a warmer dust temperature ($40<T_{d}<80$ K, \cite{sand,hughes} and references 
therein). The warmer dust could be due to the additional source of UV radiation
provided by the hidden quasar present in the nucleus of $IRAS$10214+4724 
\cite{goodrich}.

If we apply the relation between the star formation rate and the
$FIR$ luminosity : $SFR = \Psi 10^{-10} L_{FIR}/L_{\odot} \ M_{\odot}yr^{-1}$,
where $\Psi=0.8-2.1$ (\cite{rr2} and references therein), and we adopt 
$\Psi=1.5$, we derive $SFR \sim 570-2250 \ h_{50}^{-2} \ M_{\odot}yr^{-1}$,
which is larger than the typical $SFRs$ of low-$z$ ULIGs ($\approx$10--100 
$M_{\odot}yr^{-1}$ \cite{sand}), but comparable to that of $IRAS$10214+4724 
($\approx$660 $M_{\odot}yr^{-1}$ \cite{hughes}). If HR10 contains an obscured 
AGN heating the dust, then only a fraction of $L_{FIR}$ can be ascribed to 
star formation. The H$\alpha$ line detected in a very low resolution
spectrum of HR10 \cite{graham} has a signal-to-noise ratio insufficient 
to establish the presence of a broad component of this line, which
would imply that the nucleus of the quasar is not severely obscured.
However, the $K$-band morphology of HR10 is spiral-like and not strongly 
nucleated, and it does not support the presence of a directly visible quasar 
nucleus.

Due to the steep rise of the grey-body dust spectra towards $\lambda_{rest} 
\sim 100-200 \mu$m, the observations at longer wavelengths make the sub-mm 
and mm flux (at a fixed observed $\lambda$) of a dusty galaxy roughly 
constant for approximately $1<z<10$ \cite{fra,blain,hughes}.
For instance, we moved HR10 to three different redshifts ($z=0.7,2.2,
3.3$) and we computed how its properties would change accordingly 
adopting $\beta=1.5$ and an optically thin grey-body : $S_{\nu} 
\propto \nu^{\beta} B_{\nu}(T)$. We find that the inferred dust temperature 
would increase from $\approx$15-20 K at $z=0.7$ to $\approx$ 45-50 K 
at $z=3.3$. $L_{FIR}$ and $M_{d}$ would be $\sim 0.1,7,25 \times10^{13}$ 
$h_{50}^{-2}$ L$_{\odot}$ and $\sim 30,10,5.6\times10^{8}$ $h_{50}^
{-2}$ M$_{\odot}$ for $z=0.7$ ($T_d=18$ K), $z=2.2$ ($T_d=35$ K),
and $z=3.3$ ($T_d=45$ K) respectively. The main implication is that
a galaxy with {\it observed} (sub)-mm properties as those of HR10 would be 
classified a dusty ultra-luminous infrared galaxy in a wide range
of redshifts. 

The discovery of dust associated with HR10 has several implications. 
First of all, it sheds light on the nature of this galaxy, demonstrating
that HR10 is a very dusty star-forming galaxy where, similarly to 
low-$z$ ULIGs, most of the energy is emitted in the far-infrared.
The second implication is generally related to the problem of estimating 
the star formation rate in distant galaxies. The global history of the
star formation in the Universe is presently inferred mostly by the
continuum UV luminosity of optically-selected high-$z$ galaxies 
\cite{madau}, from which the derived $SFRs$ are in the range of 4-25
$h_{50}^{-2}$ M$_{\odot}$yr$^{-1}$ \cite{steid}. However, such 
estimates can be severely hampered by the presence of dust extinction 
which modifies the shape and reduces the flux of the UV spectra. 
Observations of low-redshift galaxies have also pointed out the limits 
of the UV--blue light as a $SFR$ estimator (see for instance \cite{gallagher} 
and references therein). HR10 allow us to see an extreme example of this 
problem applied to a high-$z$ system. In fact, the H$\alpha$ luminosity 
of HR10 would imply a $SFR\sim80 h_{50}^{-2}$ M$_{\odot}$yr$^{-1}$ 
\cite{graham}, whereas the luminosity of the UV continuum at 2800~\AA~ 
would suggest $SFR\sim 1 h_{50}^{-2}$ M$_{\odot}$yr$ ^{-1}$ \cite{madau}, 
both at least one order of magnitude less than the $SFR$ suggested by the 
$FIR$ luminosity. In this regard, our results show that the global star 
formation history of the Universe can be fully traced only if the effects 
of dust are taken into account. Objects like HR10 would have not been found 
by neither optical imaging surveys based on the Lyman-continuum 
break \cite{steid} or on strong emission lines, nor by $IRAS$ surveys, 
nor by traditional quasar surveys. Instead, our results demonstrate 
that the combination of optical and near-IR deep imaging, coupled 
with (sub)-mm observations, is an efficient method to find dusty galaxies 
at high-$z$. Recent deep $SCUBA$ imaging suggests indeed the presence 
of a population of faint sub-mm sources \cite{smail}. 

The observations of HR10 suggest that the star formation in distant 
objects occurs with different modes, and that the most vigorous episodes 
of star formation probably arise in dusty environments as predicted by 
several models \cite{fra,zepf,zepf2}. However, our data cannot tell us 
if HR10 is forming its first generation of stars, or if the starburst 
is occurring in an already formed system. Nevertheless, in both cases we 
are witnessing a major episode of star formation: if the burst lasts 
about $10^{7}$ years (a typical time scale for a starburst \cite{sand}), 
the total mass of gas converted into stars is of the order of $10^{10}$ 
M$_{\odot}$, which is approximately 10\% of the total mass of 
a present-day massive galaxy. 

Finally, we find relevant to note that the observed properties of HR10
fit into the predictions of \cite{guid} that the sources at $z\sim0.5-2.5$
which contribute to the cosmic $FIR-mm$ background should have fluxes 
around 10-100 mJy at $\lambda_{obs}=200\mu$m. For $18\leq T_{d}\leq45$ K, the 
expected flux of HR10 at $\lambda_{obs}=200\mu$m is in the range of 10-40 mJy. 
Moreover, we also note that the sky surface density of
dusty galaxies at $\lambda_{obs}=175\mu$m predicted by \cite{guid}
($\approx 0.05$ arcmin$^{-2}$) is within a factor of five similar
to that of the extremely red galaxies \cite{hu,thompson}.
Future observations will provide clues on what fraction of the 
extremely red galaxies contributes to the $FIR-mm$ background.

{\bf Acknowledgments.} We are grateful to Sofia Randich, David Hughes, 
Rob Ivison, Matt Lehnert and Jo Baker for useful discussions, to Esther 
Hu for checking the astrometry of HR10 and for helpful suggestions. 
We also thank the three anonymous referees for their useful and constructive 
comments.

\centerline{\bf Figure Caption}

{\bf Figure 1:} 
Filled circles: fluxes of HR10 taken from \cite{hu,graham,ivi,
jones} and this work. The $IRAS$ $3\sigma$ upper limits at
$\lambda_{obs}=12,25,60,100\mu$m were derived 
from co-added survey data with the SCANPI procedure available at 
IPAC (Infrared Processing \& Analysis Center). 
%
%
The curves show two extreme cases of grey-bodies consistent
with the data. Dashed curve : optically thin grey-body, $S_{\nu} 
\propto \nu^{\beta} B_{\nu}(T)$, with $\beta=1.5$ and $T_{d}$=18 K.
Continuous curve : grey-body optically thin at $\nu<\nu_0$=750 GHz, 
$S_{\nu} \propto B_{\nu}(T)\{1-exp[-(\nu/\nu_0)^{\beta}]\}$ \cite{hughes},
with $\beta=1.5$ and $T_{d}=45$ K. HR10 was observed for 64 minutes 
(on source) with the MPIfR 19-channel bolometer \cite{kreysa}
(1250$\mu$m, $11 ^{\prime \prime}$ (FWHM) beam, $32^{\prime \prime}$ 
chop throw, zenith opacity 0.09-0.3). Flux calibration was achieved  
using Uranus as primary calibrator and Mars and pointing quasars as 
secondary ones. The mean atmospheric level was found averaging the 
measurements of 16 channels and subtracted from the central pixel, 
which targeted HR10. The final flux at 1250$\mu$m is $4.9 \pm 0.7$ mJy. 
The simultaneous observations at 450 and 850$\mu$m were made with the 
$SCUBA$ bolometer array (Robson et al., in preparation) at the JCMT 
in photometry mode (108 minutes on source, zenith opacity 0.3 at 850$\mu$m). 
The sky background was removed from the flatfielded and despiked data by 
subtracting the average signal from 5 neighbour bolometers 
(containing no source signal). The data were calibrated using photometry 
and beam maps of Mars and Uranus. The flux at 850$\mu$m and 450$\mu$m 
resulted to be 8.7 $\pm$ 1.6 mJy and $<$180 mJy (3$\sigma$) respectively.
The $1\sigma$ error bars of the 850$\mu$m and 1250$\mu$m flux densities
indicate the statistical uncertainties, whereas the uncertainty
on the absolute flux calibration is estimated to be $\sim$15-20\% both 
in the $IRAM$ and $SCUBA$ data.

\vfil
\eject

\psfig{figure=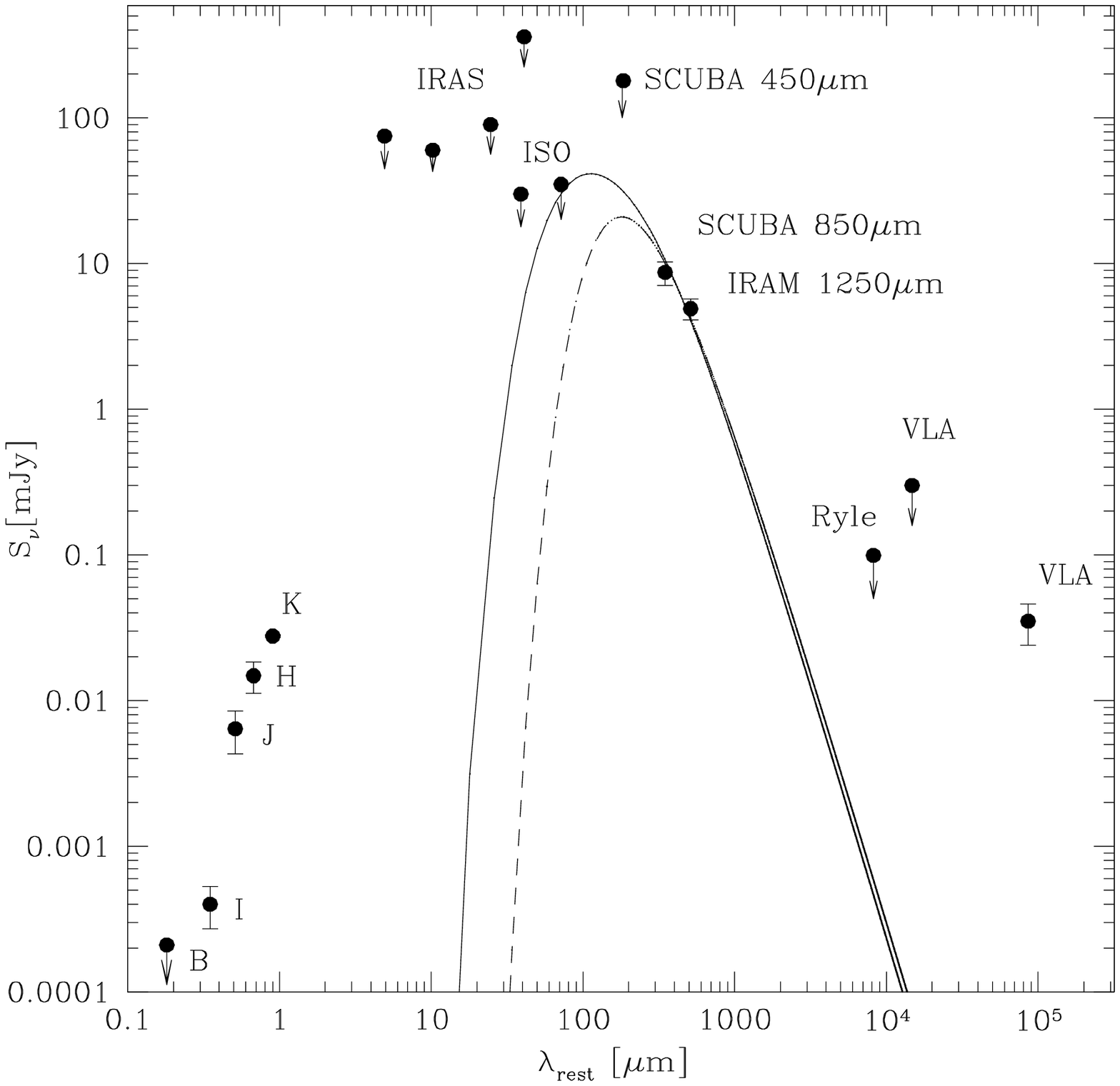,width=20cm} 

\end{document}